\documentclass[showpacs,aps,twocolumn,superscriptaddress,prb,amsmath]
{revtex4}
\usepackage{txfonts}
\usepackage{color}
\usepackage[latin9]{inputenc}
\usepackage{amssymb}
\usepackage{graphicx}
\usepackage{dcolumn}
\usepackage{bm}
\usepackage{hyperref}
\usepackage{color}

\begin{document}
\preprint{APS/123-QED}
\title{Stone-Wales graphene: A Two Dimensional Carbon Semi-Metal with Magic Stability}
\author{HengChuang Yin$^\sharp$}
\affiliation{Hunan Key Laboratory for Micro-Nano Energy Materials and Devices, Xiangtan University, Hunan 411105, P. R. China;}
\affiliation{School of Physics and Optoelectronics, Xiangtan University, Xiangtan 411105, China.}
\author{Xizhi Shi$^\sharp$}
\affiliation{Hunan Key Laboratory for Micro-Nano Energy Materials and Devices, Xiangtan University, Hunan 411105, P. R. China;}
\affiliation{School of Physics and Optoelectronics, Xiangtan University, Xiangtan 411105, China.}
\author{Chaoyu He}
\email{hechaoyu@xtu.edu.cn}\affiliation{Hunan Key Laboratory for Micro-Nano Energy Materials and Devices, Xiangtan University, Hunan 411105, P. R. China;}
\affiliation{School of Physics and Optoelectronics, Xiangtan University, Xiangtan 411105, China.}
\author{Miguel Martinez-Canales}\email{miguel.martinez@ed.ac.uk}
\affiliation{SUPA, School of Physics and Astronomy $\&$ CSEC, University of Edinburgh, Peter Guthrie Tait Road, Edinburgh EH9 3FD, United Kingdom}
\author{Jin Li}
\affiliation{Hunan Key Laboratory for Micro-Nano Energy Materials and Devices, Xiangtan University, Hunan 411105, P. R. China;}
\affiliation{School of Physics and Optoelectronics, Xiangtan University, Xiangtan 411105, China.}
\author{Chris J. Pickard}
\email{cjp20@cam.ac.uk}
\affiliation{Department of Materials Science $\&$ Metallurgy, University of Cambridge, 27 Charles Babbage Road, Cambridge CB3$~$0FS, United Kingdom}
\affiliation{Advanced Institute for Materials Research, Tohoku University 2-1-1 Katahira, Aoba, Sendai, 980-8577, Japan}
\author{Chao Tang}  
\affiliation{Hunan Key Laboratory for Micro-Nano Energy Materials and Devices, Xiangtan University, Hunan 411105, P. R. China;}
\affiliation{School of Physics and Optoelectronics, Xiangtan University, Xiangtan 411105, China.}
\author{Tao Ouyang}
\affiliation{Hunan Key Laboratory for Micro-Nano Energy Materials and Devices, Xiangtan University, Hunan 411105, P. R. China;}
\affiliation{School of Physics and Optoelectronics, Xiangtan University, Xiangtan 411105, China.}
\author{Chunxiao Zhang}
\affiliation{Hunan Key Laboratory for Micro-Nano Energy Materials and Devices, Xiangtan University, Hunan 411105, P. R. China;}
\affiliation{School of Physics and Optoelectronics, Xiangtan University, Xiangtan 411105, China.}
\author{Jianxin Zhong}
\affiliation{Hunan Key Laboratory for Micro-Nano Energy Materials and Devices, Xiangtan University, Hunan 411105, P. R. China;}
\affiliation{School of Physics and Optoelectronics, Xiangtan University, Xiangtan 411105, China.}
\date{\today}

%
%
\begin{abstract}
A two-dimensional carbon allotrope, Stone-Wales graphene, is identified in stochastic
group and graph constrained searches and systematically investigated by first-principles
calculations.
Stone-Wales graphene consists of well-arranged Stone-Wales defects, and it can be
constructed through a 90$^\circ$ bond-rotation in a $\sqrt{8}$$\times$$\sqrt{8}$ super-cell
of graphene.
Its calculated energy relative to graphene, +149 meV/atom, makes it more stable than
the most competitive previously suggested graphene allotropes
We find that Stone-Wales graphene based on a $\sqrt{8}$ super-cell is more stable than those
based on $\sqrt{9} \times \sqrt{9}$, $\sqrt{12} \times \sqrt{12}$ and
$\sqrt{13} \times \sqrt{13}$ super-cells, and is a ``magic size"
that can be further understood through a simple ``energy splitting and inversion" model.
The calculated vibrational properties and molecular dynamics of SW-graphene confirm that it is dynamically stable.
The electronic structure shows SW-graphene is a semimetal with distorted, strongly anisotropic Dirac cones.
\end{abstract}
\maketitle

Elemental carbon adopts many different allotropes, including the three-dimensional (3D) cubic diamond, hexagonal diamond and graphite,
the two-dimensional (2D) graphene \cite{1} and graphynes \cite{2}, the one-dimensional (1D) nanotubes \cite{3}, as well as the zero-dimensional (0D) fullerenes \cite{4}, due to its ability to hybridize in sp, sp$^2$ and sp$^3$. Low-dimensional carbon materials, such as graphene and nanotubes, have attracted much scientific interest in view of their novel electronic and mechanical properties \cite{1, 5, 6}. In particular, graphene is
a well-known Dirac-cone material that exhibits high carrier mobility \cite{1, 5, 6} and the quantum Hall effect \cite{5, 7, 8} due to its
semi-metallicity contributed from the active $\pi$-electrons \cite{8, 9}. The successes of isolating graphene \cite{1, 10, 11} from graphite
have spurred many efforts in searching for other 2D carbon allotropes beyond graphene. Graphynes \cite{12, 13} are alternative ways
to fill the 2D space with mixed sp-sp$^2$ carbon atoms. They contribute rich electronic properties, including
semiconductors, semi-metals and metals. Although graphynes with sp-hybridized bonding are energetically less stable than the
pure sp$^2$-hybridized graphene, graphite and nanotubes, there are a few graphynes that have been experimentally synthesized \cite{2, 14}.

The honeycomb-like graphene is not the only way to topologically fill 2D space with sp$^2$ hybridized carbon. Many other hypothetical graphene allotropes have been previously proposed.
For example, the pentaheptites (R$_{57-1}$ and R$_{57-2}$) containing exclusively
5-7 rings proposed in 1996 \cite{15} and 2000 \cite{16, 17}, the Haeckelite sheets (H$_{567}$ and O$_{567}$) proposed by Terrones \cite{17},
the low-energy  $\psi$-graphene previously proposed by Cs\'anyi \cite{18} and recently investigated by Li \cite{19}, the dimerites
(dimerite I, II and III with metallic property) and octites (metallic octite M$_1$, M$_2$, M$_3$ and semiconducting octite SC) constructed
through defects patterns \cite{20,21,22,23}, the biphenylene sheets (New-W and New-C) containing 4-6-8 rings \cite{24, 25}, the T-graphene
containing 4-8 rings \cite{26,27,28}, the PO-graphene containing 5-8 rings (OPG-L and OPG-Z) \cite{29, 30}, the graphene superlattice structures
(including pza-C$_{10}$) \cite{31}, the 187-C$_{65}$, 191-C$_{63}$ and 127-C$_{41}$ sheets \cite{32}, the 5-6-8 rings HOP-graphene \cite{33},
$\delta$-graphene \cite{34} and pentahexoctite \cite{35}, as well as the recently proposed phagraphene \cite{36}. Most of these 2D
carbon allotropes are more energetically stable than the experimentally viable graphynes \cite{2, 14} and C$_{60}$ \cite{4}. In particular,
the most stable $\psi$-graphene \cite{8} and phagraphene \cite{36} possess energies just about 165 meV/atom and 201 meV/atom higher than that
of graphene, and might be synthesized in future experiments.

\begin{figure}[hbt]
   \center
   \includegraphics[width=\columnwidth]{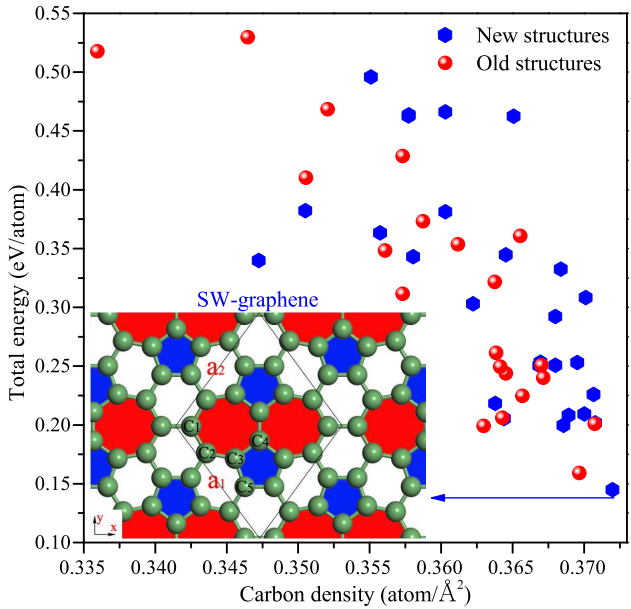}
   \caption{Calculated energies (relative to graphene) and carbon densities of
    the new discovered (blue hexagon) and previously proposed (red balls) graphene allotropes with energy relative to graphene less than 550 meV per atom. \emph{Inset:} The crystalline view of the most stable one (SW-graphene) with symmetry of $Cmmm$, showing its topological pattern containing five-, six- and seven-fold carbon rings.
    %
    }\label{fig1}
\end{figure}


Previously suggested 2D carbon allotropes, other than octite SC \cite{23}, have
less than 20 atoms per cell. In this work, we have performed stochastic group and
graph-constrained searches for 2D carbon allotropes with up to 24 atoms per unit cell.
Graph-constrained searches have been performed using our recently developed RG$^2$ code \cite{40}.
Structural details, energetics and properties have all been computed
using density functional theory (DFT) and the PBE exchange-correlation functional \cite{44}
as implemented in Vienna Ab-initio Simulation Package (VASP) \cite{41}.
Carbon was modelled with a $2s^2\ 2p^2$ PAW potential \cite{42,43} and the energy cutoff
was set to be 500~eV. The Brillouin Zone (BZ) sampling meshes were denser than 0.21 \AA$^{-1}$
to ensure the convergence. The vibrational spectrum of SW-graphene was computed using the phonopy code \cite{45}.
\begin{figure}
   \center
   \includegraphics[width=\columnwidth]{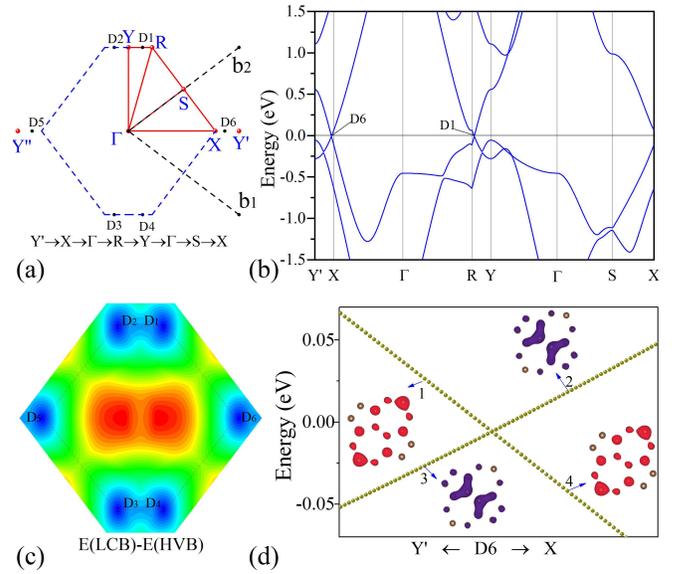}
   \caption{The first Brillouin zone (BZ) (a), the band structures (b), the 2D color mapping
   of the energy differences between LCB and HVB (c), as well as the band decomposed charge
   density distributions of SW-graphene (d).}\label{fig2}
\end{figure}

Our RG$^2$ searches  produced nearly all the previously proposed 2D carbon allotropes.
It further identified 33 previously unreported competitive structures as shown in Fig.~ref{fig1}, including the most stable allotrope known to date. Its structure contains well arranged Stone-Wales defects (SWD, 5577-ring segments)\cite{37,38,39}
on a $\sqrt{8}$ graphene supercell, and we call it Stone-Wales graphene (SW-graphene hereafter).
SW-graphene, inserted in Fig.~\ref{fig1}, has 16 carbon atoms and one SWD in its primitive unit cell. The SWDs periodically connect to each other with six-fold carbon rings boundary. As indicated in Fig.~\ref{fig1}, the primitive cell (with optimized lattice constants of a=b=6.683 $\AA$, c=12.00 $\AA$ and $\gamma$ =105.779$^\circ$) of SW-graphene contains 5 inequivalent carbon atoms C$_1$, C$_2$, C$_3$, C$_4$ and C$_5$, located at positions 4h (0.09, 0.09, 0.5), 8q (0.299, 0.062, 0.5), 8q (0.497, 0.211, 0.5), 4j (0.564, 0.435, 0.5) and 4j (0.682, 0.143, 0.5), respectively. The bond lengths (1.378-1.458 $\AA$) and bond angles (106.11-133.95$^\circ$) \cite{SM} in SW-graphene are very close to those in ideal graphene, which indicate that it should be a low-lying 2D carbon allotrope.

We calculated the energies of the new discovered and previously proposed graphene allotropes \cite{SM}. Those with energy relative to graphene less than 550 meV/atom are shown in Fig.~\ref{fig1}. The most stable three graphene allotropes previously proposed, $\psi$-graphene \cite{18, 19}, octite M$_1$ \cite{22} and phagraphene \cite{36}, lie 159, 199 and 201~meV/atom above graphene, respectively. The relative energy of
SW-graphene is +149~meV/atom, which indicates that it is energetically more favorable than either. We notice that the carbon density of SW-graphene (0.372 atom/\AA$^2$) is also comparable to that of graphene (0.379 atom/\AA$^2$), larger than all the previously proposed 2D carbons.

With its remarkable energetic stability, SW-graphene might be synthesized if it is dynamically stable. The phonon dispersion bands and
phonon density of states of SW-graphene show no imaginary frequency mode in the whole Brillouin Zone \cite{SM}, indicating SW-graphene is dynamically stable. \emph{Ab initio} NVT molecular dynamics on a $3\times 3$ supercell and a timestep of 1~fs confirm that SW-graphene is
dynamically stable at 300~K for at least 5~ps. No structural changes occur, with the only distortion being thermal oscillations of atoms near
their equilibrium positions (Fig.~S3 in the supplemental material \cite{SM}). That is to say, SW-graphene should be stable at room temperature.

As discussed in previous literatures concerning the synthesis of $\psi$-graphene \cite{18} and Phagraphene \cite{36}, defect topology \cite{39, 50}
and molecular assembly \cite{46, 51} are two potential routes to synthesize SW-graphene. We notice that SW-graphene can be decomposed as ethylene
and benzene (or cyclopentene and propylene) as indicated in Fig.~S3. That is to say, catalyzer-associated assembly of ethylene and benzene
(or cyclopentene and propylene) on a proper substrate could be a way to synthesize SW-graphene.

Most of the previously proposed graphene allotropes, such as $\psi$-graphene \cite{18, 19} and octite M$_1$ \cite{22} are metallic.
A few of them have been reported as semiconductors and semimetals, such as octite SC \cite{23} and Phagraphene \cite{36}. Our calculations
show that SW-graphene is a semi-metal with distorted Dirac-cones. There are two equivalent Dirac-cones in its first Brillouin zone (BZ)
as indicated in Fig.~2 (a). The calculated band structure of SW-graphene along the pathway of Y'--X--$\Gamma$--R--Y--$\Gamma$--S--X is shown
in Fig.~2 (b). A symmetry analysis shows that, along Y'--$\Gamma$, at least one band must cross the Fermi energy (D$_6$). We can see that there
is a Dirac-point located at the point (D$_1$) between R--Y. The calculated density of states (DOS) at Fermi-level is zero, which confirms
that such a Dirac-point is not formed by band-folding. Fig.~2 (c) shows the energy difference between the highest valance band (HVB) and the
lowest conduction band (LCB). We can see that there are six zero points (Dirac-points) within that region. D$_2$ is equivalent to D$_1$ due to
mirror symmetry along $y$-axis. D$_6$ (D$_5$) is equivalent to D$_2$ (D$_1$) via a reciprocal lattice vector shift of b$_1$ (negative b$_2$) as indicated in Fig.~2 (a).
D$_4$ and D$_3$ are equivalent to D$_1$ and D$_2$, respectively, due to the mirror symmetry on $x$-axis.

To further understand the formation of
the Dirac-cones in SW-graphene, the band-decomposed charge density distributions of the four points (1, 2, 3, 4) near the Dirac-point D$_6$ are
investigated. As shown in Fig.~2 (d), we can see that the charge densities of points 1 and 4 are very similar to each other. Those for points 2
and 3 also appear in same distribution. This indicates that the Dirac-points in SW-graphene are formed by band-cross similar to those in phagraphene \cite{36}.

To show the cone-like feature of these Dirac-points, the 3D surface band structures of the HVB and LCB near D6 are plotted.
As inserted in Fig.~3, the HVB and LCB cross each other at point D$_6$ and form a distorted Dirac-cone with obvious anisotropicity.
Based on the calculated surface band structures, we then investigated the direction-dependent Fermi velocities of the Fermions in
cone D$_6$ as shown in Fig.~3. The corresponding Fermi velocities are 
given by
$\upsilon$$_f$ = E(k)/$\hbar$$|k|$. Our calculated Fermi velocities of Phagraphene
along $x$, $y$ and -$y$ directions are 6.54$\times$10$^5$ m/s, 6.34$\times$10$^5$ m/s and 4.17$\times$10$^5$ m/s, respectively.
The largest and smallest Fermi velocities in graphene are 8.26$\times$10$^5$ m/s and 7.74$\times$10$^5$ m/s, respectively.
These results are consistent with those reported before \cite{36}. However, we find that the largest Fermi velocity
in phagraphene is not the 6.54$\times$10$^5$ m/s along the $x$ direction. It is about 7.02$\times$10$^5$ m/s along
the $\pi$/6 direction, which was not reported in previous literature \cite{36}.

The result in Fig.~3 shows that the Fermi velocities of SW-graphene
along the $x$ ($\theta=0^\circ$), -$x$ ($\theta=180^\circ$) and $y$ ($\theta=90^\circ$) directions are 7.78$\times$10$^5$ m/s,
5.16$\times$10$^5$ m/s and 4.25$\times$10$^5$ m/s, respectively. Fermions in SW-graphene along the $x$ direction possesses the largest
velocity of 7.78$\times$10$^5$ m/s. Along the direction of $\theta$$\approx$7$\pi$/12, the Fermi velocity (4.09$\times$10$^5$ m/s) is the smallest.
These results indicate that SW-graphene possesses anisotropic electronic properties similar to phagraphene \cite{36},
which are more obvious than those in graphene.

\begin{figure}
\center
\includegraphics[width=\columnwidth]{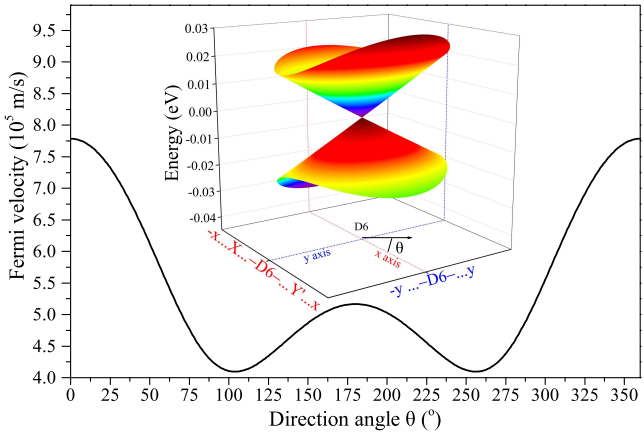}
\caption{The 3D plot of the Dirac-cone formed by the highest valence band (HVB) and the lowest conduction band (LCB) in the
vicinity of the Dirac point (D$_6$) and the corresponding direction-dependent Fermi velocities for Fermions in SW-graphene.}\label{fig3}
\end{figure}

\begin{figure}
    \center
    \includegraphics[width=\columnwidth]{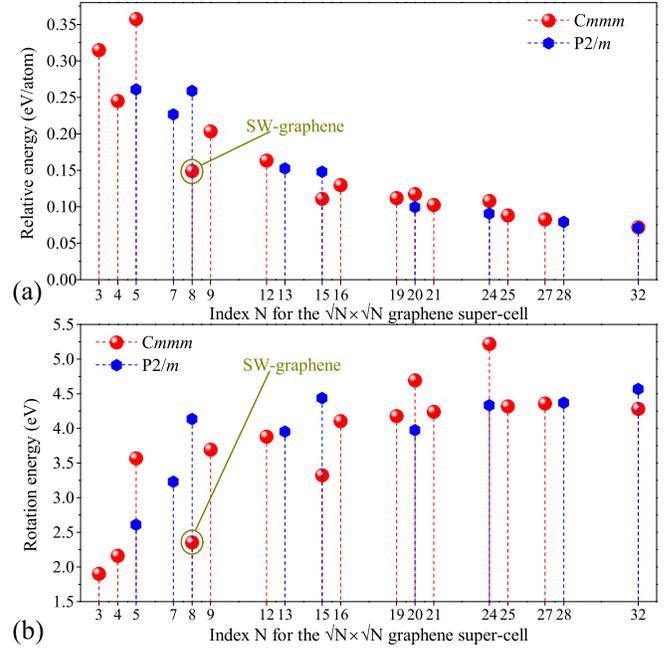}
    \caption{Relative energies \emph{(a)} and SWD rotation energies
      ($E_{\sqrt{N}} - N E_\textrm{graphene}$) (b)
      of $\sqrt{N} \times \sqrt{N}$ graphene super-cells with a single SWD.
      SW-graphene corresponds to $N=8$.
      Red balls denote structures that adopt $Cmmm$ symmetry; blue hexagons denote
      structures adopting $P2/m$ symmetry.
      }
      \label{fig4}
\end{figure}

\begin{figure*}[t]
\center
\includegraphics[width=\textwidth]{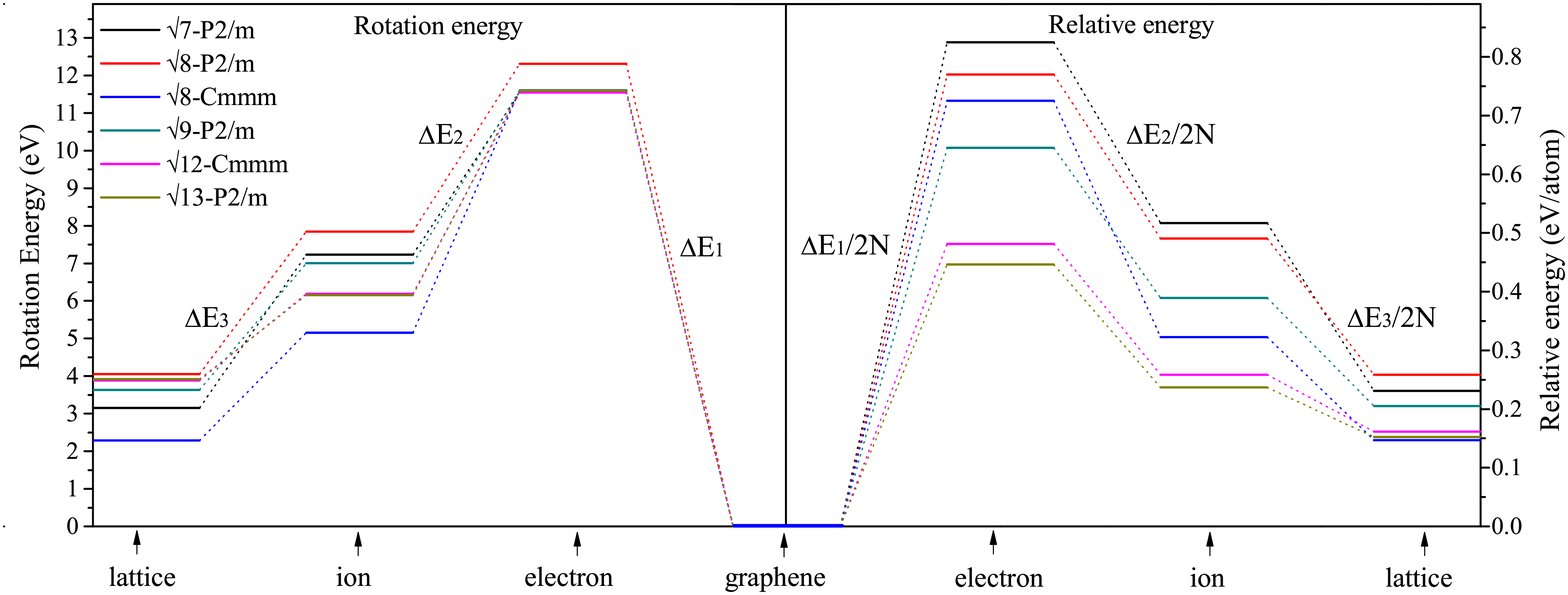}
\caption{Energy splitting and inversion model (left: rotation energies eV per cell; right: relative energies eV/atom relative to graphene) based on the comparison of energy changes between $\sqrt{7}$, $\sqrt{8}$-$Cmmm$, $\sqrt{8}$-$P2/m$, $\sqrt{9}$,
$\sqrt{12}$ and $\sqrt{13}$ in three steps after bond-rotation, namely, electronic minimization in fixed configuration (electron),
ionic relaxation (ion), as well as full lattice and ionic relaxation (lattice).}\label{fig5}
\end{figure*}

We have also computed the band structure using maximally-localized Wannier functions using
Quantum-Espresso \cite{48} and Wannier90 \cite{49}. Our starting guess has been p$_z$ orbitals
on each atom, sp$^2$ orbitals on alternating atoms where possible, and $\sigma$-bonds on the
remaining bonds. This replicates the DFT bands up to at least 4 eV above the Fermi energy \cite{SM}.
Projecting into orbitals show a picture similar to graphene: the low-lying
bands come from sp$^2$ orbitals, while the bands near Fermi energy can be described by p$_z$ interactions.
The result of such a projection is shown in Fig.~S6. This opens up the creation of an effective Hamiltonian
for SW-graphene, but even a nearest-neighbour model has 15 symmetry-independent hopping parameters. The Berry
phase around a Dirac cone takes a nontrivial value of $\pi$.

SW-graphene can be understood as a simple 90$^\circ$ carbon bond rotation \cite{SM}.
However, we show now the stability of SW-graphene is not just the result
of a single defect in a supercell, and that SW-graphene is a structure in its own right.
Fig.~\ref{fig4}~(a) shows the energies of $\sqrt{N}\times\sqrt{N}$ graphene
supercells with a single SWD. Larger cells are more stable, as expected,
despite relaxation in smaller cells reducing a large fraction of the rotation
energy penalty of a SWD. However, SW-graphene ($\sqrt{8}$-$Cmmm$) is 50~meV/atom more stable than $N=9$ and,
unexpectedly, more stable than $N<15$. The rotation energy in SW-graphene is almost as low as for the $N = 3, 4$
cases. This suggests $N=8$ is a \emph{``magic size''}, followed by other two magic sizes of $\sqrt{15}$ and $\sqrt{20}$.

A simple ``energy splitting and inversion'' model is used to understand
the ``magic stability'' of SW-graphene as shown in Fig.~\ref{fig5}.
After bond-rotation, we compare the energies and rotation energies of
$\sqrt{7}$, $\sqrt{8}$-$Cmmm$, $\sqrt{8}$-$P2/m$, $\sqrt{9}$, $\sqrt{12}$
and $\sqrt{13}$ in three steps: electronic minimization in fixed configuration,
ionic relaxation, as well as full lattice and ionic relaxation.
In fixed configuration, other than an obvious energy penalty for
$\sqrt{8}$-$P2/m$, other candidates show roughly the same $\Delta E_1$.
The averaged values $\Delta$E$_1$/2N split or reverse the energy positions of
$\sqrt{7}$, $\sqrt{8}$-$Cmmm$, $\sqrt{8}$-$P2/m$, $\sqrt{9}$, $\sqrt{12}$ and $\sqrt{13}$.
After bond-rotation, the defected atoms \cite{dratio} in the system adjust
their position to minimize the energy. With larger ratio of defected
atoms, ionic and lattice relaxations release more energy
$\Delta E_2$ and $\Delta E_3$ in small cells ($\sqrt{7}$,
$\sqrt{8}$-$Cmmm$, $\sqrt{8}$-$P2/m$) than larger systems
($\sqrt{9}$, $\sqrt{12}$ and $\sqrt{13}$).
The results in Fig.~\ref{fig5} show how $\sqrt{8}$-$Cmmm$ releases the largest amount of
energy on ionic relaxation, which is eventually the cause of the stability
of SW-graphene.

We finally check if Dirac-cones are common features in the
SW-graphene allotropes family \cite{SM}.
SW-graphene allotropes $\sqrt{3}$, $\sqrt{4}$, $\sqrt{5}$-$Cmmm$,
$\sqrt{7}$, $\sqrt{9}$, $\sqrt{12}$ and $\sqrt{15}$-$Cmmm$
are all metallic.
Those named as $\sqrt{5}$-$P2/m$, $\sqrt{8}$-$P2/m$, $\sqrt{13}$, $\sqrt{15}$-$P2/m$, $\sqrt{19}$,
$\sqrt{20}$-$Cmmm$,$\sqrt{20}$-$P2/m$, $\sqrt{21}$, $\sqrt{24}$-$P2/m$, $\sqrt{28}$ and $\sqrt{32}$-$P2/m$,
are semiconductors with direct or indirect band gaps.
Only a few of them are semi-metals featuring Dirac-cones:
$\sqrt{8}$-$Cmmm$ (SW-graphene), $\sqrt{16}$, $\sqrt{24}$-$Cmmm$,
$\sqrt{25}$, $\sqrt{27}$ and $\sqrt{32}$-$Cmmm$.

In summary, a 2D carbon allotrope is discovered using our recently developed RG$^2$ code.
It is named SW-graphene in view of its structural feature consisting of the well-known Stone-Wales defects.
The calculated total energy shows that SW-graphene is energetically more favorable than the previously proposed
phagraphene and $\psi$-graphene. It is also confirmed to be a dynamically stable carbon phase according to its
phonon dispersion and molecular dynamics. The calculated band structures shows that SW-graphene is a semimetal with distorted
Dirac-cones, containing Fermions with direction-dependent Fermi velocities. In view of its remarkable stability and
novel properties, SW-graphene is a desirable target for future synthesis. A family of SW-graphene allotropes constructed through
90$^\circ$-rotation of carbon-bonds in given $\sqrt{\mathrm{N}}$$\times$$\sqrt{\mathrm{N}}$ graphene super-cells are investigated.
We find that $\sqrt{8}$-$Cmmm$ SW-graphene is an interesting system with ``magic stability" and not all the SW-graphene allotropes
are semi-metallic like $\sqrt{8}$-$Cmmm$ SW-graphene. The ``magic stability" of $\sqrt{8}$-$Cmmm$ SW-graphene is further discussed based on
a theoretical ``energy splitting and inversion" model.

This work is supported by the National Natural Science Foundation of China (Grants No. 11704319, No. 11647063, No. A040204, and No. 11204261), the National Basic Research Program of China (No. 2015CB921103), the Natural Science Foundation of Hunan Province, China (Grant No. 2016JJ3118), and the Program for Changjiang Scholars and Innovative Research Team in University (No. IRT13093). MMC is grateful for computational support from the UK Materials and Molecular Modelling Hub, which is partially funded by EPSRC (EP/P020194), for which access was obtained via the UKCP consortium and funded by EPSRC grant EP/P022561/1. C. J. P. is supported by the Royal Society through a Royal Society Wolfson Research Merit award.
\bibliographystyle{apsrev}

\begin{thebibliography}{99}
\bibitem{dratio}Defected atoms are the selected two atoms (bond) and their four surroundings. The ratio of defected atoms is defined
as (2+4)/2N, in which 2 and 4 denotes the selected two atoms and the four surroundings, respectively. 2N is the total atoms number in the cell.
For example, the ratio for SW-graphene with size of $\sqrt{8}$ and $\sqrt{32}$ are 6/16 and 6/64, respectively.
\end{thebibliography}

\end{document}